\documentclass[aps,jmp,amsmath,amssymb,preprint,12pt]{revtex4-1}
\usepackage{dcolumn}
\usepackage{dcolumn}
\usepackage{bm}
\usepackage{graphicx}
\usepackage{subfigure}
\usepackage{physics}
\usepackage{subscript}
\usepackage{soul}
\usepackage[utf8x]{inputenc} 
\usepackage{verbatim}
\usepackage[usenames,dvipsnames]{color}

\begin{document}

\title{Tuning the lattice thermal conductivity in Bismuth Telluride through Cr-doping}
\author{Ajit Jena$^{1}$}
\author{Seung Cheol Lee$^1$}
\email{seungcheol.lee@ikst.res.in}
\author{Satadeep Bhattacharjee$^{1}$}
\email{satadeep.bhattacharjee@ikst.res.in}

\affiliation{$^{1}$Indo Korea Science and Technology Center, Bangalore-560065, India}

\begin{abstract}Decreasing thermal conductivity of a thermoelectric material is always a prerequisite for its potential application. Using first-principle calculations, we examine the magnetism induced change in lattice thermal transport in bismuth telluride. The source of magnetic moment, Cr in the doped system, weakly magnetizes the coordinated Te atoms to make the latter's phonon softer than that in the pure compound. Though the transition metal dopants do not participate directly in the heat conduction process, the anharmonicity induced by them favor in reducing the lattice thermal conductivity. Large anharmonicity in $(Bi_{0.67}Cr_{0.33})_2Te_3$ reduces the in-plane room temperature lattice thermal conductivity by $\sim 79\%$. The thermal conductivity, strictly, does not vary monotonically with doping concentration. Even, for any particular doping level, the thermal conductivity is different for different configurations which is related to the internal energy of the system. We found that the internal energy variance of 0.03 $eV$ would reduce the in-plane thermal conductivity of the room temperature lattice by at least 60$\%$ for 50$\%$ doping.
\end{abstract}

\maketitle
\section{Introduction}
Designing new high-performance thermoelectric materials or enhancing existing efficiency will undoubtedly attract considerable attention in the context of power generation or refrigeration. The efficiency of a thermoelectric material is assessed by the electrical power factor ($S^2\sigma$) and thermal conductivity ($\kappa$), where $\kappa$ includes contributions both from electrons and lattice ($\kappa=\kappa_{el}+\kappa_{ph}$), $S$ being the Seebeck coefficient and $\sigma$ is the electrical conductivity. A good thermoelectric material should possess high $S^2\sigma$ and low $\kappa$. Furthermore, materials, in which $\kappa_{el}$ is negligible and $\kappa$ is mainly contributed by the phonons, will be potential candidates to yield improved thermoelectric performance. One of the best thermoelectric materials is bismuth telluride, $Bi_2Te_3$ (BT) which has intrinsically low $\kappa_{ph}$ and has been studied extensively in bulk,\cite{goldsmid1, goldsmid2} alloys,\cite{alloys1, alloys2} thin films, nanostructures,\cite{nano-str} and by band engineering\cite{band-eng} or by chemical doping.\cite{in-dop, pb-dop, mi-dop}

When BT is doped with In, the system behaves as electron donors and as a result it reduces both electrical resistivity and $\kappa_{ph}$.\cite{in-dop} Similarly, metal and Iodide co-doping effect enhances the electrical conductivity and decreases the lattice heat conduction of BT.\cite{mi-dop} Reduced $\kappa_{ph}$ is also observed in metallic Pb-doped and PbTe-precipitated BT.\cite{pb-dop} The reduced lattice thermal conductivity in all these foreign element doped BT is mainly attributed to the mechanisms such as formation of point defects, lattice distortion, grain boundary phonon scattering etc. There are numerous reports on transition metal (TM) doped BT family where the interest is strongly biased towards the properties of diluted magnetic semiconductors \cite{tc177, pssb, larson, jo, lee} and the heat conduction in these systems is negligibly explored. Very recently, it has been suggested that spin-phonon coupling in determining the spin and thermal transport properties in magnetic insulators and topological magnetic systems is an important feature of physics.\cite{spin-phonon1} In fact, it is essential to understand the working stability of spin-phonon coupled high performance thermal magnet, where thermal transport has a pivotal role.\cite{spin-phonon2} In an experimental work, it is shown that the lattice thermal conductivity in Cr-doped BT decreases due to point defect scattering and the overall performance increases by $25\%$.\cite{vaney} However, they show that the change in the thermal conductivity is not monotonic with the doping level. This has encouraged us to further investigate the spin induced phonon transport in $(Bi_{1-x}Cr_x)_2Te_3$ using first-principles tool. To the best of our knowledge, our attempt is first of its kind in the TM doped BT family of compounds.

The lattice thermal conductivity is described by the specific heat ($c_{ph}$), group velocity and mean free path ($l_{ph}$) of the phonons. In the low temperature limit, below the Debye temperature ($\Theta_{D}$), the phonons are scattered by the impurities making $l_{ph}$ temperature independent and $\kappa_{ph}$ varies as $c_{ph}$ ($\propto T^3$). When $T > \Theta_{D}$, $c_{ph}$ abides by the Dulong-Petit law where phonon-phonon (p-p) anharmonic scattering is dominated and $l_{ph}$ varies as $1/T$ implying $\kappa_{ph} \propto 1/T$. Further in materials like BT, Umklapp process is known to dominate the p-p anharmonic scattering and $\kappa_{ph}$ varies as $T^{-1}$.\cite{qiu, park} We calculate $\kappa_{ph}$ tensor by solving phonon Boltzmann transport equation (BTE) \cite{shengbte} using Eq.~\ref{eq:kph}. We show that though the TM do not take part directly in the heat conduction process, they magnetize the neighboring Tes and induce anharmonicity which in turn reduces the thermal conductivity. Larger doping levels do not proportionately reduce the thermal conductivity. Due to difference in internal energy among various configurations of any doping level, the phonon transport is different. Maximum reduction occurs in the configuration with larger internal energy. We feel that our results make an important fundamental contribution to understanding and tailoring spin induced thermal conductivity in tetradymitic structure compounds ($Bi_2Te_3$, $Bi_2Se_3$, $Sb_2Te_3$), in particular and TM doped other non-magnetic thermoelectric materials in general.

\section{Computational Methodology}

\begin{figure}
\centering
\includegraphics[width = 7cm]{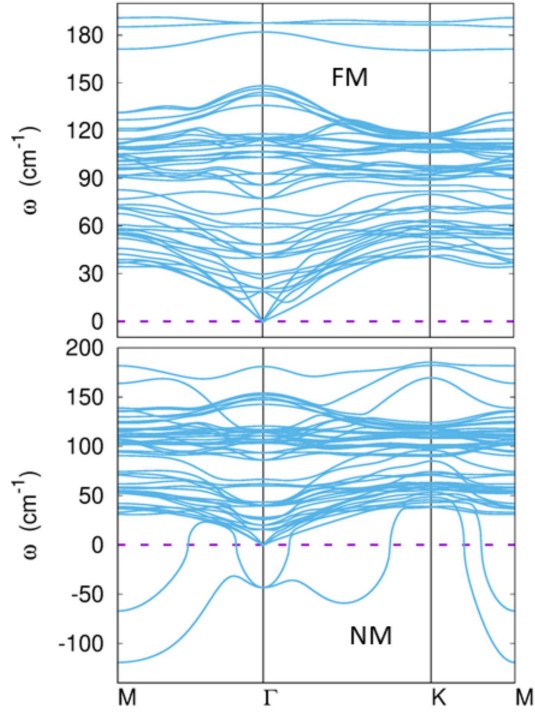}
\caption{Ferromagnetic (FM) and non-magnetic (NM) phonon dispersion relations of $(Bi_{0.83}Cr_{0.17})_2Te_3$ along the high symmetry-lines of hexagonal lattice. The lattice is unstable in the NM phase.}
\label{fig:fm-nm_phdis}
\end{figure}

\begin{figure}
\centering
\includegraphics[scale = 0.25]{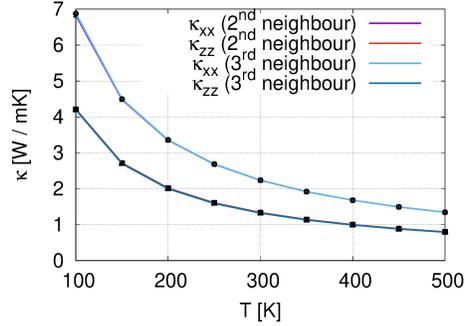}
\caption{Lattice thermal conductivity of $Bi_2Te_3$ for third order force constants calculated up to second and third nearest-neighbors interaction.} 
\label{fig:cutoff}
\end{figure}

Boltzmann Transport Equation has been proven to be successful in describing the heat transport in insulators, semiconductors and metals.\cite{allen, lindsay1, lindsay2, shengbte, epw} We use ShengBTE package \cite{shengbte} to calculate $\kappa_{ph}$ tensor by supplying second and third order inter atomic force constants. The $\kappa_{ph}$ as a function of temperature \textit{T} is calculated as  

\begin{equation}
\kappa_{ph}^{\alpha\beta}= \frac{1}{k_BT^2NV}\sum_\lambda (\hbar\omega_\lambda)^2f_\lambda^0(1+f_\lambda^0)v_\lambda^\alpha F_\lambda^\beta,
\label{eq:kph}
\end{equation}

where $k_B$, \textit{N}, \textit{V}, $\lambda$, $\hbar$, $\omega$, $f_\lambda^0$ and $v$ respectively denote the Boltzmann constant, number of \textit{q} points, volume of the unit cell, phonon mode, reduced Plank's constant, angular frequency, Bose-Einstein distribution function and the phonon group velocity. $F_\lambda = \tau_\lambda^0 (v_\lambda + \Delta_\lambda)$ is the projection of mean free displacement with $\tau_\lambda^0$ being the relaxation time of phonon mode $\lambda$ and $\Delta_\lambda$ stands for the correction term, having dimension same as velocity, removes the deviation predicted by relaxation time approximation.

We employ pseudo-potential based density-functional theory (DFT) and density-functional perturbation theory (DFPT) as implemented in Quantum ESPRESSO \cite{giannozzi} within the framework of local density approximation (LDA) to compute the force constants. Norm-conserving pseudo-potential is used in the calculations and the kinetic energy cutoff for the planewave is taken as 60 Ry. The electronic integration over the Brillouin zone is approximated by the Gaussian smearing of 0.001 Ry for the self-consistent calculations. The k-grid of $16\times16\times1$ and q-grid of $4\times4\times1$ are considered respectively for the self-consistent and second order force constant calculations. The third order force constant calculation is performed on supercells containing $3\times3\times1$ unit cells including up to third nearest-neighbors interaction. Finally, all the necessary inputs are provided to ShengBTE code to calculate $\kappa_{ph}$ on a q-grid of $32\times32\times1$.

To generate unpaired spins and thereby non-zero net magnetic moment in BT we have substituted the TM element Cr at Bi sites which are reported to be energetically more favorable sites than Te.\cite{larson, fav-site} Minimal doping concentration of 3\textit{d} TM elements makes BT a diluted ferromagnetic semiconductor.\cite{pssb, jo, lee} Since our aim is to limit the phonon conductivity by spin-lattice coupling we have considered larger doping concentration $x = 0.17, 0.33, 0.5$ and $0.67$ in $(Bi_{1-x}Cr_x)_2Te_3$. To do so, hexagonal conventional unit cell of bismuth telluride containing 15 atoms has been used in the calculations. This should allow us to obtain a qualitative understanding of the spin effect on the phonon transport in BT. Another reason for considering larger doping levels is to reduce the computational cost in the phonon calculations. In Fig.~\ref{fig:fm-nm_phdis}, we have presented the phonon dispersion of $(Bi_{0.83}Cr_{0.17})_2Te_3$ along the high symmetry lines of hexagonal lattice obtained from the magnetic and non-magnetic (NM) calculations. The presence of large negative frequency in the NM dispersion curve suggests that the lattice is unstable in NM phase and therefore the same is not considered any further in the present study. It is worth mentioning that the calculated thermal conductivities should converge with respect to cutoff distance (nearest neighbor) that is involved in the third order force constant calculations. As we can seen in Fig.~\ref{fig:cutoff} our result is well converged for two different nearest neighbors. Therefore, to minimize the number of SCF calculations in doped systems, up to second nearest neighbors is taken into account for the computation of third order force constant. For example, it requires 736 more SCFs in case of $(Bi_{0.83}Cr_{0.17})_2Te_3$ when we move from second (cutoff $=3.89$ \AA{}) to third (cutoff $=4.47$ \AA{}) nearest neighbors. The number of additional SCFs are more for larger doping levels. Isotope scattering is also taken into account in calculating $\kappa_{ph}$. However, its effect on $\kappa_{ph}$ is negligible.

\section{Results and Discussions}

\begin{figure}
\centering
\includegraphics[scale = 0.35]{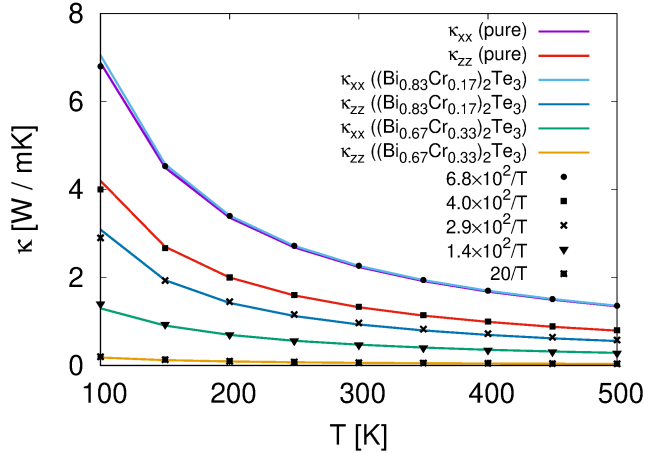}
\caption{Temperature dependent lattice thermal conductivity of $(Bi_{1-x}Cr_x)_2Te_3$ ($x$ = 0.0, 0.17 and 0.33) along in-plane and out-of-plane directions. The thermal conductivities are fitted with the function $1/T$.} 
\label{fig:kappa-pure_dop}
\end{figure}

\begin{figure}
\centering
\includegraphics[scale = 0.7]{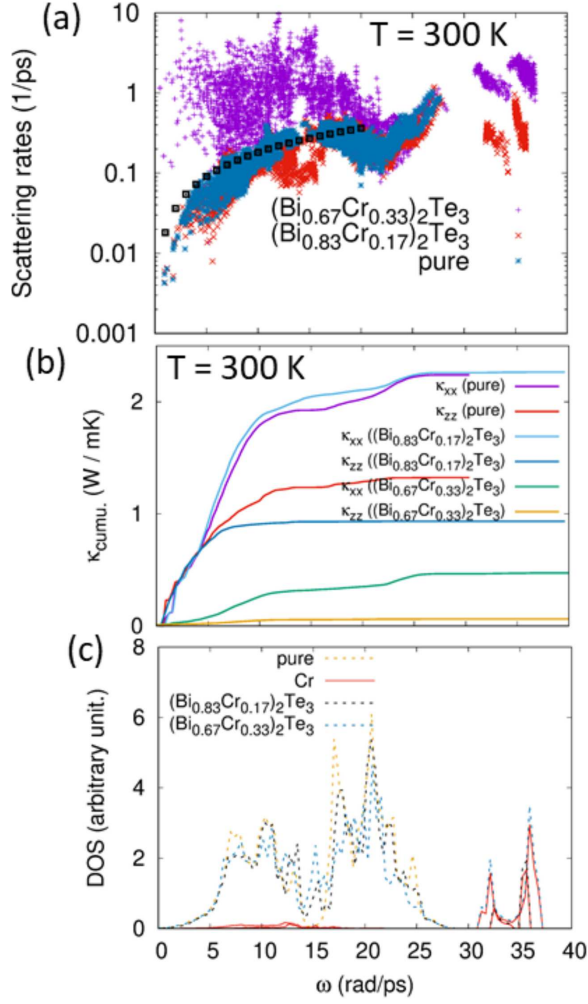}
\caption{Room temperature (a) phonon-phonon anharmonic scattering rates and (b) cumulative thermal conductivity of pure and Cr-doped bismuth telluride varies with phonon frequency. The black squares in (a) represents the linear fit. (c) Total phonon density of states of pure and Cr-doped bismuth telluride. As it is marked in the figure the higher energy phonons are solely contributed by the Cr dopant but contribute less in the thermal transport.}
\label{fig:sr-cum_kap-phdos}
\end{figure}

\begin{figure}
\centering
\includegraphics[height = 6cm]{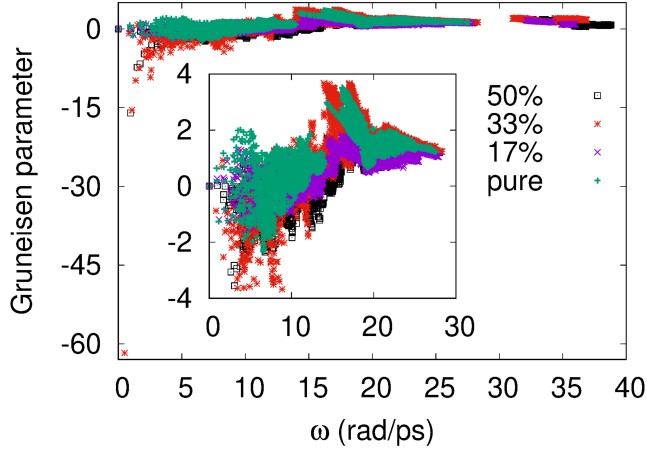}
\caption{Gr\"uneisen parameter in $(Bi_{1-x}Cr_x)_2Te_3$ ($x$ = 0.0, 0.17, 0.33 and 0.5) as a function of phonon frequency. Substantial anharmonicity occurs in case of $x$ = 0.33. The inset displays the pattern in the smaller range.} 
\label{fig:wvsgp}
\end{figure}

\begin{figure*}
\center
\includegraphics[scale = 0.55]{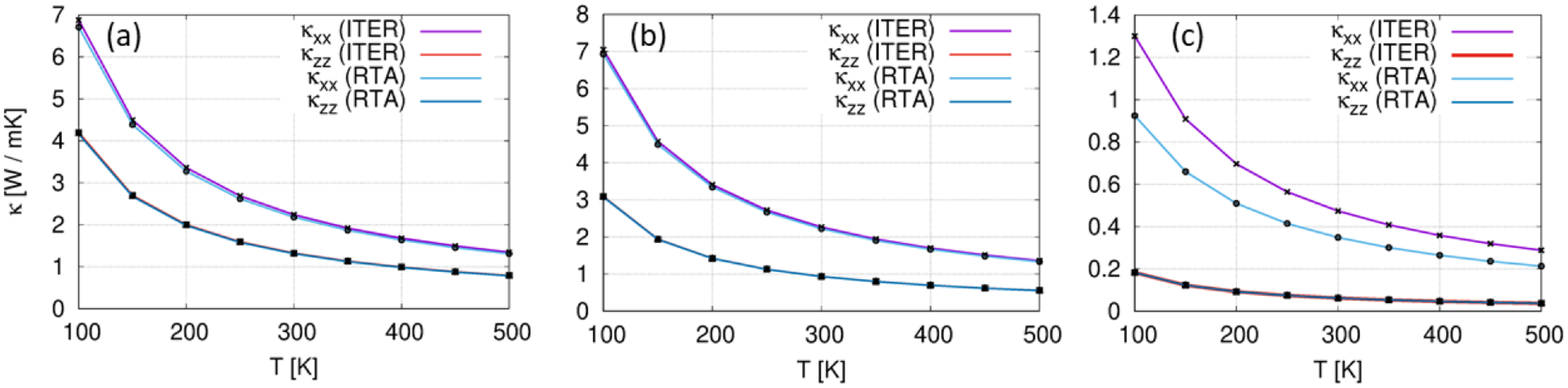}
\caption{ Lattice thermal conductivities of (a) $Bi_2Te_3$ (b) $(Bi_{0.83}Cr_{0.17})_2Te_3$ and (c) $(Bi_{0.67}Cr_{0.33})_2Te_3$ calculated using iterative approach and relaxation time approximation. The agreement between the two suggests that the scattering mechanism is dominated by Umklapp process.}
\label{fig:iter-rta}
\end{figure*}

The lattice heat transport in bismuth telluride is anisotropic due to the difference in phonon group velocities along in-plane and out-of-plane directions.\cite{qiu, 4cbroido, park} As a result of this, the lattice thermal conductivity in out-of-plane direction is lesser ($\sim 40\%$) than in the in-plane direction.\cite{qiu, 4cbroido, park} Moreover, the acoustic phonons are the major contributors to the phonon heat conduction and Umklapp process is found to dominate over the normal three-phonon process.\cite{qiu, park} In Fig.~\ref{fig:kappa-pure_dop} we compare the temperature dependent lattice thermal conductivities of pure and doped (17$\%$ and 33$\%$) bismuth telluride. Larger doping levels significantly affects the heat transport both in in-pane and out-of-plane directions. For example, we find that the in-plane (out-of-plane) lattice thermal conductivity is decreased by $\sim 79\%$ ($ > 90\%$) in case of $(Bi_{0.67}Cr_{0.33})_2Te_3$. There are two factors that can influence $\kappa_{ph}$ upon doping, the phonon group velocity and the p-p anharmonic scattering. As expected, the phonon group velocity along in-plane direction is larger compared to the out-of-plane one (ESI).$^\dag$ However, the velocity for pure and doped cases are nearly same for each component. And from the room temperature anharmonic scattering rates in Fig.~\ref{fig:sr-cum_kap-phdos}(a), it can be seen that due to reduced lifetime of phonons in $(Bi_{0.67}Cr_{0.33})_2Te_3$ the thermal conductivity is decreased significantly. The behavior of the scattering rates in all 3 cases is clearly reflected on the cumulative $\kappa$ plot in Fig.~\ref{fig:sr-cum_kap-phdos}(b). Owing to the increased point defects in $(Bi_{0.67}Cr_{0.33})_2Te_3$ the pattern of its scattering rates are largely different from the other two. The substantial enhancement of p-p anharmonic scattering in $x$ = 0.33 case is the outcome of large anharmonicity in the system. The anharmonicity is described in Fig.~\ref{fig:wvsgp} through the estimation of Gr\"uneisen parameter. The p-p anharmonic scattering exceeds the boundary scattering in $(Bi_{0.67}Cr_{0.33})_2Te_3$.$^\dag$ The boundary scattering in BT is typically strong.\cite{park} Further, from the phonon density of states in Fig.~\ref{fig:sr-cum_kap-phdos}(c) we learn that the higher energy phonons are solely contributed by the Cr dopant. However, phonons having energy $>$ 30 $rad/ps$ do not contribute anything to the heat transport (see Fig.~\ref{fig:sr-cum_kap-phdos}(b)) and hence stand as trivial in the present case. The lattice thermal conductivity is majorly contributed by the phonons with frequency below 20 $rad/ps$.

Further, to validate our results, we have shown the calculated thermal conductivities using iterative and relaxation time approximations in Fig.~\ref{fig:iter-rta}. The greater agreement between the results of two approaches is an indication of the scattering mechanism dominated by Umklapp process.\cite{shengbte} We note that the original in-plane Umklapp scattering process in $(Bi_{0.67}Cr_{0.33})_2Te_3$ is modified which can be seen from the disagreement between the $\kappa_{xx}$ values. This is an artefact and this anomaly is caused by the minimal negative phonon frequency (-2.30 $cm^{-1}$) of in-plane acoustic mode at the zone center. We have also quantified the contribution of acoustic phonons on $\kappa_{ph}$ and found that the acoustic modes contribute up to $74\%$ and $79\%$ of lattice thermal conductivity respectively along in-plane and cross-plane directions at $300 K$ for the pure system. Also in $(Bi_{0.83}Cr_{0.17})_2Te_3$, we get larger contributions from the acoustic phonon branches ($79\%$ along in-plane and $96\%$ along out-of-plane). However, there is a deviation in case of $(Bi_{0.67}Cr_{0.33})_2Te_3$ along the in-plane direction ($50\%$ along in-plane and $75\%$ along out-of-plane). This deviation can be related to the case that causes the earlier anomaly and can also be understood from the missing scattering processes as shown in Fig.~\ref{fig:sr-cum_kap-phdos}(a)).
 
\begin{figure}
\centering
\includegraphics[scale = 0.55]{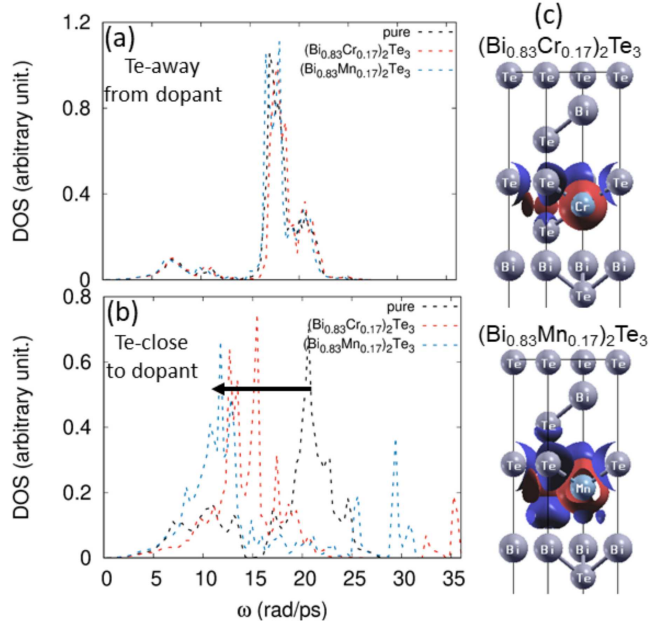}
\caption{Partial phonon density of states (DOS) of Te situated (a) away from the magnetic dopant and (b) close to the magnetic dopant. (c) Spin density plots of $(Bi_{0.83}Cr_{0.17})_2Te_3$ (up) and $(Bi_{0.83}Mn_{0.17})_2Te_3$ (down) are presented in order to demonstrate the effect of induced opposite spin polarization. The Te phonon-DOS in $(Bi_{0.83}Mn_{0.17})_2Te_3$ are found to lie lower (see in (b)) in energy while comparing with $(Bi_{0.83}Cr_{0.17})_2Te_3$. The backward arrow represents lowering of the Te-phonon frequency. The $\kappa \sim \omega$ relation which is used by the ShengBTE code \cite{shengbte} is given in Eq.~\ref{eq:kph}.}
\label{fig:p-phdos}
\end{figure}

The defect site as well as its surroundings should be the subject of interest, since together they constitute the disorder in the system. We notice that the non-magnetic neighboring Te atoms are weakly magnetized by the magnetic dopant due to induced opposite spin polarization (IOSP). The IOSP is demonstrated in Fig.~\ref{fig:p-phdos}(c) by the spin density plots. In this process they seem to be heavier than the Tes which are away from the magnetic dopant. The partial phonon density of states of Te which are sitting both near and away from the magnetic dopant are compared in Fig.~\ref{fig:p-phdos}(a,b). The frequency of the weakly magnetized Te shifts downward and thereby reduces the thermal conductivity in the doped system. Though it is not shown here, the number of weakly magnetized Te atoms are more in larger doping level. Similar IOSP effect is also seen when we consider another 3\textit{d} TM element (Mn). In case of $(Bi_{0.83}Mn_{0.17})_2Te_3$, we find relatively larger down shift of phonon frequency. The lattice thermal conductivity is directly related to the phonon frequency and the relation that the ShengBTE code \cite{shengbte} uses to compute thermal conductivity is already given in Eq.~\ref{eq:kph}. We note that further detailed heat transport mechanism in $(Bi_{0.83}Mn_{0.17})_2Te_3$ is not included in the current study and we plan to report it separately.

\begin{figure*}
\centering
\includegraphics[height=9.55cm]{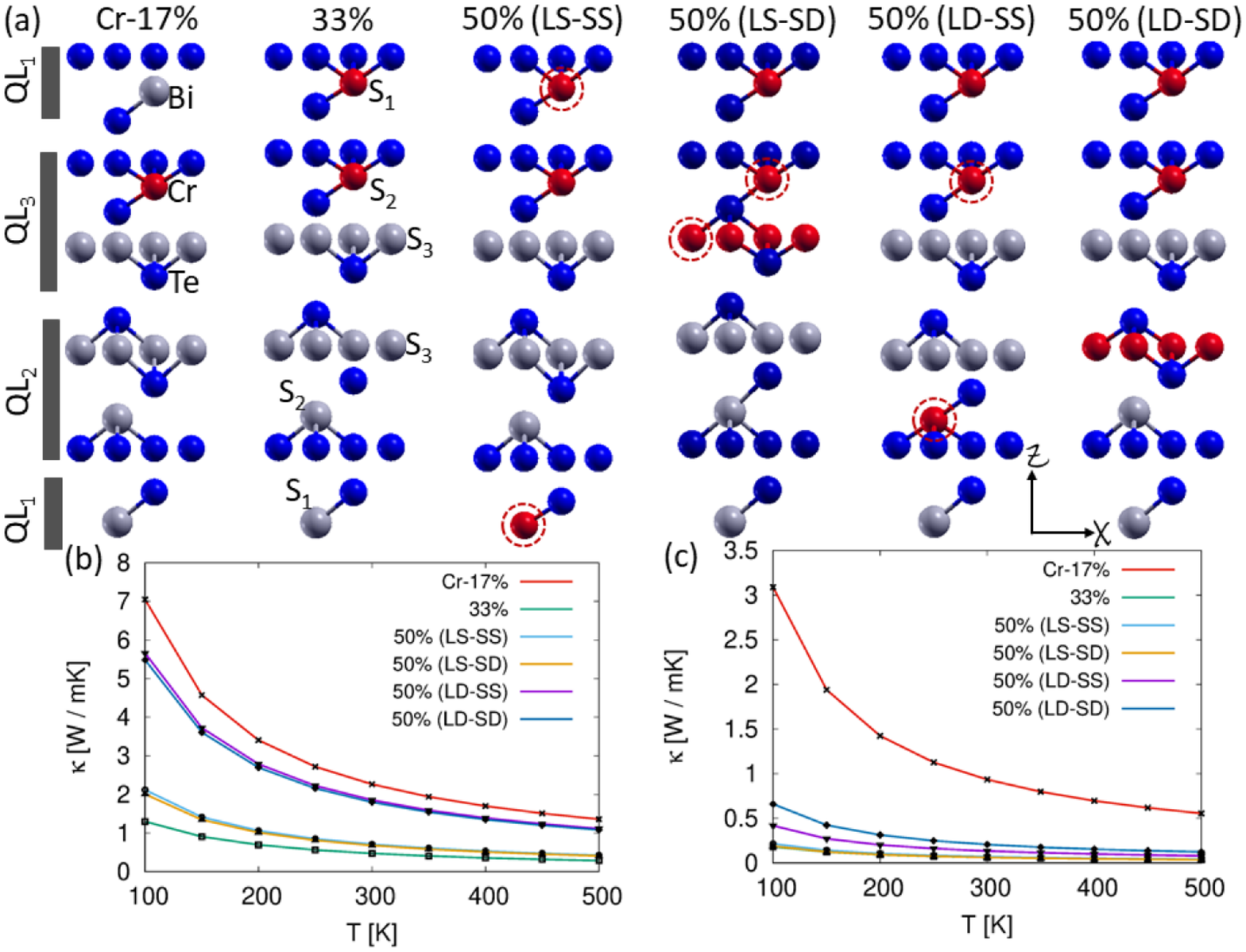}
\caption{(a) Arrangements of dopants in $(Bi_{1-x}Cr_{x})_2Te_3$ lattice for various doping percentages \textit{viz.} $17\%$, $33\%$ and four different configurations of $50\%$ considered in this paper. The three quintuple layers and three different symmetry sites on which the six Bi atoms sit are respectively named as QL$_1$-QL$_2$-QL$_3$ and S$_1$-S$_2$-S$_3$. When any two out of three dopants in 50$\%$ case are belong to either a same QL or same symmetry, they are marked by a doted circle. For example, both the QL and symmetry are same (different) in case of LS-SS (LD-SD). LS-SS, LS-SD, LD-SS and LD-SD respectively represent layer same-symmetry same, layer same-symmetry different, layer different-symmetry same and layer different-symmetry different. (b) In-plane and (c) out-of-plane lattice thermal conductivities of all the configurations (shown in (a)) are compared.}
\label{fig:all-dop}
\end{figure*}

Another interesting feature of BT is its crystal structure, the unit cell has three quintuple layers (QLs) and three different symmetry sites on which the six Bi atoms sit.  Each QL contains two Bi and three Te atoms (one formula unit).  When we substitute one Cr in place of a Bi in any of the QL the material is $17\%$ doped.  However, for larger doping levels (e.g.  substituting two Cr atoms) both the Cr can be present in a single QL or in two different QLs. In case of $50\%$ doping, all the three dopants can be inside at least two QLs or in three different QLs. Similar is the case for three different symmetry sites where every individual symmetry can occupy two Bi. However, each QL does not necessarily belong to any specific symmetry and vice-versa. As we can see from Fig.~\ref{fig:all-dop}(a), both the Bi in only QL$_1$ belong to the same symmetry S$_1$. Though from the doping level view point all possible configurations look same, the internal energy is different. More comprehensive methods to find all possible configurations by permutation for any doping level may be used. We study few of them (Fig.~\ref{fig:all-dop}(a)) considering the presence of dopant in the QL and symmetry site. The three QLs and symmetries are named as QL$_1$-QL$_2$-QL$_3$ and S$_1$-S$_2$-S$_3$ respectively. We note that the features of the partial phonon density of states of any pair of Bi atoms that belong to the same symmetry are exactly same and among the pairs it differs a little.  

\begin{figure}
\centering
\includegraphics[scale = 0.61]{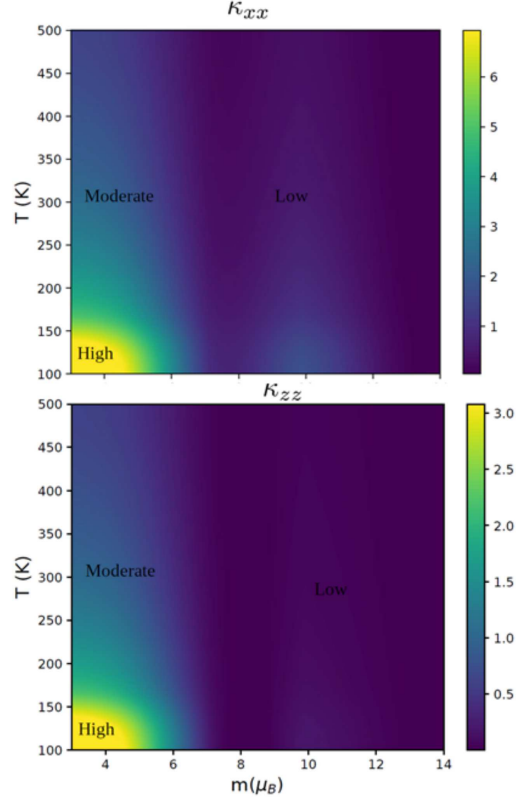}
\caption{Phase diagram displaying the effect of magnetic moment on the temperature dependent lattice heat transport in Cr-doped bismuth telluride.} 
\label{fig:pd}
\end{figure}

It is important to examine the phonon transport in different configurations considering the arrangements of dopants in different QLs and symmetry sites. In Fig.~\ref{fig:all-dop}(b,c), we  have compared in-plane and out-of-plane thermal conductivities of  $17\%$ and $33\%$  doping levels (one configuration from each level) with four different  configurations for $50\%$ doping level. The values for $50\%$ doping are larger than that of $33\%$ doping and they are even different among different configurations considered here. The smaller $\kappa$ values in $33\%$ doping case are attributed to the occurrence of larger anharmonicity (Fig.~\ref{fig:wvsgp}) in the system. This incoherent behavior of $\kappa$ with respect to the doping level can be due to the magnetic fluctuations which may be suppressed by the applied field.\cite{spin-phonon1} Earlier, non-monotonic nature of lattice thermal conductivity in Cr-doped bismuth telluride has also been observed.\cite{vaney} In Fig.~\ref{fig:all-dop}(b) for 50$\%$ doping, the relatively higher values in LD configuration (in comparison to LS one) is related to its lower internal energy. LD is more stable than LS by 0.03 $eV$. The thermal conductivity in the lower energy configuration is higher and vice-versa. LD stands for the configuration in which three dopants are arranged in three different QLs and in case of LS, three dopants are arranged in two QLs, making one of the QL common for two of the dopants. In any case the lattice thermal conductivity in Cr-doped $Bi_2Te_3$ is lesser than in the pure compound, verified after finding further lower values in $(Bi_{0.33}Cr_{0.67})_2Te_3$ (compared to $(Bi_{0.67}Cr_{0.33})_2Te_3$) which is actually a Cr-concentrated system. The $\kappa$ values of all doping levels ($x$ = 0.17, 0.33, 0.5 and 0.67) are compared in the phase diagram shown in Fig.~\ref{fig:pd}. The phase diagram displays the effect of magnetic moment on both the components of lattice $\kappa$ at all temperatures. At low temperature (within the range 100-150 K) and for small values of magnetic moment ($\leq 5 \mu_B$) the thermal conductivity is in it's maximum. As the temperature is raised, even for the small values of magnetic moment,  $\kappa$ decreases moderately due to the temperature induced anharmonic effects. However, when the magnetic moment exceeds 6.4 $\mu_B$, we see that the $\kappa$ values are dramatically changed. For both components, $\kappa$ values drop by $50-60\%$ within a broad temperature range of 100-500K.  The total energy of the system gets significantly effected due to the exchange interaction between the Cr-atoms given by the Hamiltonian, ${\cal H}=\sum_{ij}J_{ij}{\bf S}_i^{Cr}{\bf S}_j^{Cr}$. Here ${\bf S}_i^{Cr}$ is the spin of the Cr atom at site i. In comparison to the pure BT, this additional energy contribution due to the Cr-spins: $E_{mag}=<{\cal H}>=\sum_{ij}J_{ij}<{\bf S}_i^{Cr}{\bf S}_j^{Cr}>$ affects the forces on the atoms \cite{Anders,Satadeep} in a way that lowers the $\kappa$. Such magnetism induced lowering of thermal conductivity, particularly low $\kappa$ values at room temperature signifies the utility of magnetic moment to enhance the $ZT$ factor in BT.

\section{Summary and Conclusions}

In summary, carrying out \textit{ab initio} calculations, we show that the magnetic dopant (Cr) decreases the lattice thermal conductivity in $Bi_2Te_3$. High energy phonons are purely contributed by the lighter magnetic dopant and they are not directly involved in decreasing the thermal conductivity. We find that the in-pane (out-of-plane) lattice thermal conductivity is reduced by $\sim 79\%$ ($ > 90\%$) in case of $(Bi_{0.67}Cr_{0.33})_2Te_3$. Increasing doping level simply does not reduce the thermal conductivity further. The thermal conductivity reduction trend follows the magnitude of anharmonicity generated by the dopant which induces opposite spin polarization on the coordinated Te atoms. Additionally, with the cost of 0.03 $eV$ of internal energy, the room temperature in-plane lattice thermal conductivity can be decreased by at least 60$\%$ of a specific doping case ($x$ = 0.5).

\section*{Conflicts of interest}
There are no conflicts to declare.

\section*{Acknowledgements}
A. Jena acknowledges the computational and financial support from IKST.

\bibliography{rsc} 
\bibliographystyle{apsrev}

\end{document}